\documentclass[twocolumn,pra]{revtex4-1}

 \usepackage{dsfont}
 \usepackage{mathdots}
 \usepackage{amsmath}
  \usepackage{amssymb}
  \usepackage{color}
 \usepackage{amsthm}
\usepackage{graphicx}% Include figure files
\usepackage{dcolumn}% Align table columns on decimal point
\usepackage{bm}% bold math
\usepackage{color}
\usepackage{hyperref}
\newcommand{\proj}[1]{|#1\rangle\langle#1|}
\begin{document}

\title{Bounding the entanglement of N qubits with only four measurements}
\author{S. M. Hashemi Rafsanjani, C. J. Broadbent, J. H. Eberly}
\affiliation{ Rochester Theory Center and the Department of Physics
\& Astronomy\\
University of Rochester, Rochester, New York 14627}
\email{hashemi@pas.rochester.edu}
%\author{.........}
%\email{......@pas.rochester.edu}
%\author{ll}
%\author{}

%\preprint{}

\date{\today}

\begin{abstract} 
  We introduce a measure for the genuinely N-partite (all-party) entanglement of N-qubit states using the trace distance metric, and find an algebraic formula for the GHZ-diagonal states. We then use this formula to show how the all-party entanglement of experimentally produced GHZ states of an arbitrary number of qubits may be bounded with only four measurements.  
\end{abstract}


% PACS
\maketitle
     
 \newcommand{\dg}[1]{#1^{\dagger}}
 \newcommand{\reci}[1]{\frac{1}{#1}}
 \newcommand{\ket}[1]{|#1\rangle}
 \newcommand{\nim}{\frac{1}{2}}
 \newcommand{\om}{\omega}
 \newcommand{\te}{\theta}
 \newcommand{\la}{\lambda}
 \newcommand{\beqa}{\begin{eqnarray}}             %Begin Equation Array
 \newcommand{\eeqa}{\end{eqnarray}}               %End Equation Array
 \newcommand{\nn}{\nonumber}                      %No Number
 \newcommand{\bra}[1]{\langle#1\vert}                 % Bra
 \newcommand{\ipr}[2]{\left\langle#1|#2\right\rangle}
  \newcommand{\up}{\uparrow}
   \newcommand{\down}{\downarrow}
     \newcommand{\dn}{\downarrow}         % Inner Product

\section{Introduction}
Quantum mechanical systems can be used for real world applications of significant importance, including quantum information processing \cite{Guhne2009,RevModPhys.81.865}, quantum cryptography \cite{nielsen:2000}, and quantum metrology  \cite{Giovannetti19112004}. Many of these applications require entanglement between more than two parties; see, e.g., \cite{Guhne2009,RevModPhys.81.865}. Such systems also allow us to probe the transition from quantum to classical behaviors in increasingly complex systems \cite{ISI:000284832600011,PhysRevLett.106.130506}. These applications are of sufficient importance to have prompted a number of experimental initiatives for generating entanglement between many parties \cite{PhysRevLett.82.1345,Monroe:2000cw,Zhao:2004er,Leibfried-2005,PhysRevLett.103.150501,Gao:2010kx, PhysRevLett.106.130506,Yao:2012fp}. Additionally, achieving quantum systems with scalable architectures continues to motivate much research.\\

In tandem with experimental efforts to create entanglement in many-party systems there has been a theoretical effort aimed at quantifying the entanglement in many party systems. Of particular importance is the entanglement existing collectively between all $N$ parties of an $N$-party system, which we call all-party entanglement \footnote{Entanglement existing between all parties of a multiparty system is sometimes called genuinely multipartite entanglement. Here, we refer to it as the all-party entanglement to more clearly indicate that it is the entanglement held collectively by all parties of a multi-party system.}, and which plays an important role in high-precision metrology as well as other applications \cite{Giovannetti19112004,PhysRevA.85.022321,PhysRevA.85.022322}.  All-party entanglement can best be identified via its opposite, biseparability. A pure state $\ket{\psi}$ is biseparable if it has a pure reduced density matrix,  i.e., $\ket{\psi}=\ket{\psi_1}\otimes \ket{\psi_2}$ where $\ket{\psi_1}$ and $\ket{\psi_2}$ are pure states. A mixed state is biseparable if it can be written as a sum of pure biseparable states; otherwise the state is all-party entangled \cite{RevModPhys.81.865,PhysRevLett.104.210501}. 

Quantifying the all-party entanglement has proved to be a challenging task. Previous studies have produced witnesses and/or lower bounds of the all-party entanglement \cite{1367-2630-12-5-053002,PhysRevA.83.062325,PhysRevA.83.062337,PhysRevLett.106.190502,PhysRevA.88.012305,PhysRevLett.108.020502,PhysRevLett.108.230502,PhysRevA.86.062303,PhysRevA.88.012305,PhysRevLett.106.190502}. Many of these results are not well suited for use in experimental settings because they apply only for idealized noise-free states \cite{PhysRevLett.108.020502,PhysRevLett.108.230502,PhysRevA.86.062303,PhysRevA.88.012305}, require numerical methods only feasible for few-partite systems \cite{PhysRevLett.106.190502}, or require knowledge of the density operator of the state under test \cite{1367-2630-12-5-053002,PhysRevA.83.062325}. The latter requirement is a problem for multi-party systems because, in general, the number of measurements to determine the density operator scales exponentially with the number of parties \cite{Altepeter2004}. Apart from these methods, the measured value of witness operators can be used to find a lower bound for the entanglement realized in an experiment \cite{1367-2630-9-3-046,PhysRevLett.98.110502,PhysRevA.81.022307,epub27517}. Nonetheless, the efficacy of this approach for producing non-trivial lower bounds is not guaranteed, and complementary techniques are desired. In summary, quantifying all-party entanglement remains a challenge.\\ 

In this report we overcome this challenge by providing upper and lower bounds on the all-party entanglement, bounds that are experimentally relevant for any state and require only four measurements regardless of the number of qubits. To do this, we first introduce a distance measure of all-party entanglement and find an algebraic formula for its value for a special class of states called the GHZ-diagonal states. We then use this algebraic formula to derive upper and lower bounds on the all-party entanglement of any state, using only four measurements. These bounds are particularly useful for states that are close to the GHZ states. We show that for any two states $\rho,\rho^{\prime}$:
\begin{align}
\label{fidelitybound}
\mathcal{E}(\rho^{\prime})-\sqrt{1-F(\rho^{\prime},\rho)^{2}}\le\mathcal{E}(\rho)\le \sqrt{1-F(\sigma',\rho)^{2}},
\end{align}
where $\mathcal{E}(\rho)$ is the value of our entanglement measure for $\rho$, and $F(\rho^{\prime},\rho)$ is the fidelity between $\rho$ and $\rho^{\prime}$, and $\sigma'$ is the closest biseparable state to $\rho'$. When $\rho'$ is a GHZ state, the bounds in Eq.~\eqref{fidelitybound} may be determined using only four measurements. 

A specific example of the direct applicability of our result is provided by results reported recently. We use Eq.~\eqref{fidelitybound} to calculate a lower bound on the actual multi-qubit entanglement produced experimentally by \citet{PhysRevLett.106.130506}. The results  are given in Table \ref{tab:label}. \\

\begin{table}[hbt!]

  \centering 
  \begin{tabular}{@{} cccccc @{}}
 \hline \hline
    Number of ions~~~~~ & 2~~~~~&3~~~~~&4~~~~~&5~~~~~&6 \\ 
\hline
      Fidelity, \%       & 98.6~~ & 97.0~~ & 95.7~~& 94.4~~&89.2 \\
    $\mathcal{E}>$ & 0.33~~ & 0.25~~ & 0.2~~ & 0.17~~ & 0.044 \\ 
    $\mathcal{E} \%>$ & 66~~ & 50~~ & 40~~ & 34~~ & 8.8 \\
\hline \hline
  \end{tabular}
    \caption{The lower bound $\mathcal{E}$ of the entanglement of the states reported in Ref.~\cite{PhysRevLett.106.130506}. The third row is the 
    percentage of the produced entanglement lower bound relative to the entanglement of a GHZ state. }
  \label{tab:label}
\end{table}

 \section{Trace distance measure}
Our method starts by introducing the trace distance, $D(\rho,\tau)=\frac{1}{2}\text{Tr}(|\rho-\tau|)$ \cite{nielsen:2000}, to define a measure of all-party entanglement, 
\begin{align}
\mathcal{E}(\rho)=\min_{\tau \in \mathcal{BS}} D(\rho,\tau),
\end{align}
where $ \mathcal{BS} $ is the set of biseparable states, and  $|A|=\sqrt{A A^\dagger}$. The trace distance is symmetric in its arguments  and is zero if and only if its arguments are equal. The benefit of this approach, as we will show, lies in the fact that when we succeed in evaluating the measure for a given state, we can readily place tight bounds on the value of the measure for the states that are close to that state. Below we establish that $\mathcal{E}(\rho)$ is a monotone of the all-party entanglement. \\

To show that  $\mathcal{E}(\rho)$ is an entanglement monotone we prove that $\mathcal{E}(\rho)$ is convex, non-increasing under local operations and classical communication (LOCC), and invariant under local unitary transformations. We first address the
convexity of $\mathcal{E}(\rho)$, showing that $\mathcal{E}(\rho)\le \lambda_{1}  \mathcal{E}(\rho_{1})+\lambda_{2}  \mathcal{E}(\rho_{2})$ if $\rho=\lambda_{1} \rho_{1}+\lambda_{2}\rho_{2}$. If the two closest biseparable states to $\rho_{1}$ and $\rho_{2}$ are $\sigma_{1}$ and $\sigma_{2}$, respectively, then
\begin{align}
 &\mathcal{E}(\rho_{1})=D(\rho_{1},\sigma_{1}), ~~\text{and}~~ \mathcal{E}(\rho_{2})=D(\rho_{2},\sigma_{2}).
\end{align}
The convexity of the trace distance allows us to conclude that
\begin{align}\nn
&\lambda_{1} \mathcal{E}(\rho_{1})+\lambda_{2} \mathcal{E}(\rho_{2})=\lambda_{1} D(\rho_{1},\sigma_{1})+\lambda_{2} D(\rho_{2},\sigma_{2})\ge\\
&D(\lambda_{1} \rho_{1}+\lambda_{2}\rho_{2}, \lambda_{1} \sigma_{1}+\lambda_{2}\sigma_{2})=D(\rho,\sigma_{12})\ge \mathcal{E}(\rho),
\end{align}
where $\sigma_{12}=\lambda_{1} \sigma_{1}+\lambda_{2}\sigma_{2}$ is biseparable, since the convex sum of any two biseparable states is itself a biseparable state.\\

 To show that our measure is non-increasing under LOCC we use the contractive property of the trace distance  \cite{nielsen:2000}. The distance between two states cannot increase under trace-preserving quantum operations, i.e. for any two states, $\rho$ and $\rho'$
\begin{align}
D(\mathcal{M}(\rho),\mathcal{M}(\rho'))\le D(\rho,\rho')
\end{align}
where  $\mathcal{M}$ is a trace-preserving quantum operation. We note that any LOCC is a completely positive, trace-preserving map \cite{PhysRevLett.78.2275}. Now if the closest biseparable state to the state $\rho$ is called $\sigma$ then 
\begin{align}
\mathcal{E}(\rho)=D(\rho,\sigma)\ge D(\Gamma(\rho),\Gamma(\sigma))\ge \mathcal{E}(\Gamma(\rho))
\end{align}
where $\Gamma $ is an LOCC. Note that in the last inequality, since $\sigma$ is biseparable, any LOCC operation on it leads to another biseparable state.

Finally, we show that $\mathcal{E}(\rho)$ is invariant under local unitary transformations. Since the trace distance is invariant under unitary transformations \cite{nielsen:2000}, then $\mathcal{E}(\rho)=\min_{\tau \in \mathcal{BS}}D(U\rho U^\dagger,U\tau U^\dagger)$. If $U$ is a local unitary transformation then $\tau'=U\tau U^\dagger$ represents a one-to-one mapping between biseparable states, and  we can write 
\begin{align}
\mathcal{E}(\rho)=\min_{\tau'\in \mathcal{BS}}D(U\rho U^\dagger,\tau^{\prime})=\mathcal{E}(U\rho U^\dagger).
\end{align}
Thus, $ \mathcal{E}$ is invariant under local unitary transformations. This completes our proof that $\mathcal{E}$ is an entanglement monotone. 

%\section{Bounds on Entanglement}
Next we show that if the value of the entanglement measure is known for some state, one can use that value to bound the entanglement of other states. Later we will find the value of $\mathcal{E}(\rho)$ for all GHZ-diagonal states.
\begin{figure}[t]
\includegraphics[width=\columnwidth]{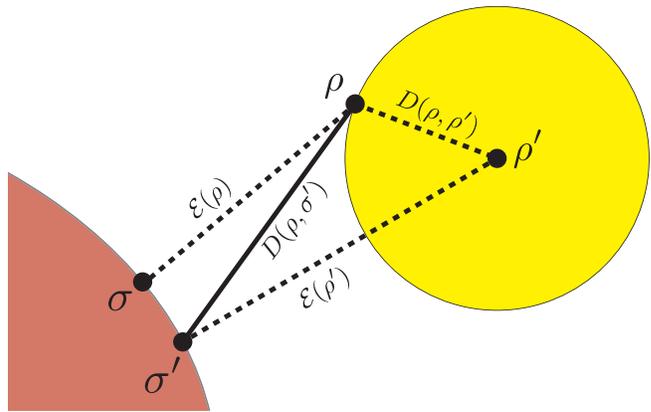}
\caption{A graphic visualization of the bounds of entanglement. The red region represents the set of biseparable states. The yellow circle represents the set of states that are closer to $\rho^{\prime}$ than $D(\rho,\rho^{\prime})$.}
\label{figure1}
\end{figure}

Let us assume we have a state $\rho$ and another state $\rho^{\prime}$ for which we know the value of $\mathcal{E}(\rho^{\prime})$. Then there exist states $\sigma$ and $\sigma'$, the closest biseparable states to $\rho$ and $\rho^{\prime}$, for which $\mathcal{E}(\rho)=D(\rho,\sigma)$ and $\mathcal{E}(\rho^{\prime})=D(\rho^{\prime},\sigma')$. Consequently we can conclude that 
\begin{align} \nn
&\mathcal{E}(\rho)= D(\rho,\sigma)\le D(\rho,\sigma')\\
 & \le D(\rho,\rho^{\prime})+D(\rho^{\prime},\sigma')=D(\rho,\rho^{\prime})+\mathcal{E}(\rho^{\prime}).
\end{align}
The last inequality is the result of the triangle inequality for our distance norm. The graphic picture in Fig.~\ref{figure1} can be helpful although our argument is purely algebraic. We can also find a lower bound for $\mathcal{E}(\rho)$ by employing the reverse triangle inequality; 
\begin{align} \nn
&\mathcal{E}(\rho)=D(\rho,\sigma) \ge D(\rho^{\prime},\sigma) - D(\rho, \rho^{\prime}) \\ 
& \ge D(\rho^{\prime},\sigma') - D(\rho, \rho^{\prime}) = \mathcal{E}(\rho^{\prime})- D(\rho, \rho^{\prime}).
\end{align}
Consequently knowledge of $\mathcal{E}(\rho^{\prime})$ allows us to bound the entanglement of $\rho$,
\begin{align}
\label{bounds}
| \mathcal{E}(\rho)-\mathcal{E}(\rho^{\prime})| \le  D(\rho, \rho^{\prime}).
\end{align}
In fact, the upper bound in \eqref{bounds} can be tightened by noting that if $\sigma'$ is known then 
\begin{align}
  \mathcal{E}(\rho)\leq D(\rho,\sigma')\leq\mathcal{E}(\rho')+D(\rho,\rho')\label{tightupper}
 \end{align}
 as shown graphically by the solid line in Fig.~\ref{figure1}. 

 Finally, using the relationship between distance and fidelity \cite{nielsen:2000},
\begin{align}
D(\rho^{\prime},\rho)\le \sqrt{1-F(\rho^{\prime},\rho)^{2}},
\end{align}
together with Eq.~\eqref{bounds} and Eq.~\eqref{tightupper}, we prove Eq.~\eqref{fidelitybound}. The above bounds will be useful provided we can find another state close enough to $\rho$ such that the bounds in Eq.~(\ref{bounds}) are non-trivial. In deriving the above bounds we have not made any assumptions about the two states $\rho$ and $\rho'$. Consequently, the value of the above bounds, and whether or not they are non-trivial, depend only on the distances (or fidelities) between states $\rho$ and $\rho'$. For example, if $D(\rho,\rho')<\mathcal{E}(\rho')$, the lower bound of entanglement becomes a non-trivial positive value. A similar argument can be made for the upper bound. Next we will find the value of our measure for GHZ-diagonal states explicitly. This allows us to find bounds on the all-party entanglement of any state using its distance to a GHZ-diagonal state.

\section{Entanglement of GHZ-diagonal states}

We will now derive the value of $\mathcal{E}(\rho')$ for the so-called GHZ-diagonal states which are mixtures of GHZ states by finding the closest biseparable state, $\sigma'$ to such states. We start with a definition and a lemma for N-qubit X-states \cite{X-state}, the class of N-qubit density matrices whose non-zero elements, in some orthonormal product basis, are diagonal and/or anti-diagonal. These matrices only have N-party coherences and tracing over any set of parties leads to a diagonal density matrix \cite{PhysRevA.82.032336,XMEMS}.\\

\textbf{Lemma}:~The closest biseparable state to an X-state is itself an X-state.\\

\textit{Proof}: If we take a density matrix $Y$ and map it to an X-state: $X=\chi(Y)$, by setting every element to zero except for the diagonal and anti-diagonal elements, it is shown in the Appendix that $\chi$ is an LOCC mapping. Let us assume the closest biseparable state to $X$, called $X_{bs}$, is not itself an X-state. The contractive property of the trace distance implies that 
\begin{align}
D(X,X_{bs})\ge D(\chi(X),\chi(X_{bs}))= D(X,\chi(X_{bs})).
\end{align}
Also note that since $X_{bs}$ is biseparable, so is $\chi(X_{bs})$. Thus if $X_{bs}$ is not itself an X-state we can find a biseparable X-state that is closer to X than $X_{bs}$. This is in contradiction to the assumption that $X_{bs}$ is the closest biseparable state to $X$. Thus we can conclude that the closest biseparable state to an X-state is itself an X-state. \\

Now we determine the closest biseparable state to the following X-state,
\begin{align}
\label{Xstate}
\hat{X}_N=\left(  \begin{array}{cccccccc}
    a_{1} & ¥ & ¥ & ¥ & ¥ & ¥ & ¥ & z_{1} \\ 
    ¥ & b_{2} & ¥  & ¥ & ¥ & ¥ & z_{2} & ¥ \\ 
    ¥ & ¥ & \ddots & ¥ & ¥ & \iddots & ¥ & ¥ \\ 
    ¥ & ¥ & ¥ & b_{n} & z_{n} & ¥ & ¥ & ¥ \\ 
    ¥ & ¥ & ¥ & z_{n}^{*} & b_{n} & ¥ & ¥ & ¥ \\ 
    ¥ & ¥ & \iddots & ¥ & ¥ & \ddots & ¥ & ¥ \\ 
    ¥ & z_{2}^{*} & ¥ & ¥ & ¥ & ¥ & b_{2} & ¥ \\ 
    z_{1}^{*} & ¥ & ¥ & ¥ & ¥ & ¥ & ¥ & b_{1} \\ 
  \end{array}
\right),
\end{align}
where $n=2^{N-1}$, and we require $|z_{i}|\le b_{i}$, $|z_1|<\sqrt{a_1 b_1}$ and $a_{1}+b_{1} +2\sum_{i} b_{i}=1$ to ensure that $\hat{X}_{N}$ is positive and normalized. We further assume that $|z_1|\ge |z_i|$. The all-party concurrence of an $N$-qubit X-matrix is given by \cite{PhysRevA.86.062303}
\begin{align}
C(\hat{X}_N)= 2\max\{0,|z_{1}|-w_{1}\},
\end{align}
where, for the X-states of the form in Eq.~(\ref{Xstate}) $w_{1}=\sum_{j\neq 1}^{n}b_{j}$. Notice that in the above X-state if $a_1=b_1$, $\hat{X}_{N}$ becomes a GHZ-diagonal state. Below, we will show that the closest biseparable state to $\hat{X}_N$ has identical elements except for $z_{1}$, which is replaced with $w_{1} z_{1}/|z_{1}|$. Consequently the value of the all-party entanglement for X-states of the above form is given by
\begin{align} \label{measureval}
\mathcal{E}(\hat{X}_N)=\max\{0,|z_{1}|-w_{1}\}.
\end{align}
In the following we further assume $z_{1} \in \mathds{R}$ and $z_{1}>0$ since we can always change the phase of the basis to achieve this, and the distance is invariant under local unitary transformations. Furthermore we assume $z_{1}=w_{1}+\epsilon>w_{1}$, otherwise $\hat{X}_{N}$ is a biseparable matrix.\\ 

Let us parameterize the closest biseparable state to $\hat{X}_{N}$ to be $\hat{\Sigma}=\hat{X}_{N}+\hat{\Delta}$;
\begin{align}
\hat{\Delta}=\left(  \begin{array}{cccccccc}
    \delta_{1}^{+} & ¥ & ¥ & ¥ & ¥ & ¥ & ¥ & \nu_{1}-\epsilon \\ 
    ¥ & \delta_{2}^{+} & ¥  & ¥ & ¥ & ¥ & \nu_2 & ¥ \\ 
    ¥ & ¥ & \ddots & ¥ & ¥ & \iddots & ¥ & ¥ \\ 
    ¥ & ¥ & ¥ & \delta_{n}^{+} & \nu_n & ¥ & ¥ & ¥ \\ 
    ¥ & ¥ & ¥ & \nu_n^* & \delta_{n}^{-} & ¥ & ¥ & ¥ \\ 
    ¥ & ¥ & \iddots & ¥ & ¥ & \ddots & ¥ & ¥ \\ 
    ¥ & \nu_2^* & ¥ & ¥ & ¥ & ¥ & \delta_{2}^{-} & ¥ \\ 
    \nu_{1}^{*}-\epsilon & ¥ & ¥ & ¥ & ¥ & ¥ & ¥ & \delta_{1}^{-} \\ 
  \end{array}
\right),
\end{align}
The X form of the difference matrix implies that $D(\hat{X}_{N},\hat{\Sigma})=||D_1||+\sum_{i>1}^{n} ||D_i||$, where $||A||=\frac{1}{2}\text{Tr}|A|$ and
\begin{align}
D_1=\left(\begin{array}{cc}
\delta_{1}^{+} & \nu_{1}-\epsilon     \\
 \nu_{1}^{*} - \epsilon & \delta_{1}^{-}    \\
\end{array}
\right), ~~~
D_i=\left(\begin{array}{cc}
 \delta_i^{+} & \nu_{i}     \\
 \nu_{i}^* & \delta_i^{-}     \\
\end{array}
\right).
\end{align}
First we turn to the $||D_1||$ contribution. Taking advantage of the triangle inequality for the eigenvalues of $D_1$ leads to
\begin{align} \nn
  ||D_1||\ge \frac{1}{2} \sqrt{(\delta_{1}^{+}-\delta_{1}^{-})^2+4|\epsilon-\nu_{1}|^2}\ge |\epsilon-\textup{Re}(\nu_{1})|.
\end{align}
Note that the above inequality implies that $\textup{Re}(\nu_{1})> 0$ since we assumed $D(\hat{X}_{N},\hat{\Sigma})<\epsilon$.
Similarly using the triangle inequality for the eigenvalues of $D_i$, we can show that $||D_i|| \ge \frac{1}{2}|\delta_{i}^{+}+\delta_i^{-}|$. Now we define $\delta_{i}=\frac{1}{2}(\delta_{i}^{+}+\delta_{i}^{-})$. We conclude
\begin{align}\nn
&\sum_{i>1} (b_i+\delta_i)\ge \sum_{i>1} \sqrt{(b_{i}+\delta_{i}^{+})(b_{i}+\delta_{i}^{-})}  \ge |w_1+\nu_{1}|\\
& \ge w_{1}+ \textup{Re}(\nu_{1}).
\end{align}
The second to last inequality arises from applying the condition of biseparability for the element $\hat{\Sigma}_{1,2n}$. Subtracting $w_{1}$ from the LHS and RHS leads to 
\begin{align}
  \sum_{i>1}||D_{i}||\ge \sum_{i>1} \delta_{i}\ge \textup{Re}(\nu_{1}).
\end{align}
Thus we conclude that $D(\hat{X}_{N},\hat{\Sigma})=||D_1||+\sum_{i>1}^{n} ||D_i|| \ge \epsilon$. One can readily check that this inequality is saturated if $\hat{\Sigma}$'s elements are all identical to $\hat{X}_{N}$ except for $z_{1}$ which is replaced with $w_{1}$. This completes our proof.

\section{Experimentally accessible bounds}

Now that we have proved Eq.~(\ref{measureval}), we need only show that if $\rho'$ is a GHZ state then $F(\rho,\rho')$ and $F(\rho,\sigma')$ can be computed with four measurements regardless the number of qubits. That $F(\rho,\rho')$ depends only on the populations and coherences $\rho_{1,1}$, $\rho_{1,2n}$, and $\rho_{2n,2n}$ and therefore requires only four measurements has been shown previously \cite{Leibfried-2005,PhysRevLett.106.130506}. From our previous analysis we find that the nearest biseparable state to the GHZ state is $\sigma'=(\proj{0\cdots0}+\proj{1\cdots1})/2$ for which it is easy to see that the fidelity, $F(\rho,\sigma')$, also only depends on the populations and coherences $\rho_{1,1}$, $\rho_{1,2n}$, and $\rho_{2n,2n}$. 

In deriving the above experimentally more accessible bounds we took advantage of the fact that the fidelity between any state and a GHZ-state depends solely on three density matrix elements, i.e. $\rho_{1,1}$, $\rho_{1,2n}$, and $\rho_{2n,2n}$. An identical argument can be made even for GHZ-states that are not of equal weighting:
\begin{align}
\ket{\psi}=\cos\theta \ket{0,0,\ldots, 0}+ \sin\theta \ket{1,1, \ldots, 1}
\end{align}
Thus the lower bound of the entanglement can be improved using a simple optimization 
\begin{align}
\mathcal{E}(\rho)\ge \max_{\theta} \left(\frac{1}{2}|\sin 2\theta|-\sqrt{1-F(\rho,\ket{\psi})^{2}} \right)
\end{align}
where the fidelity reads
\begin{align}
  F(\rho,\ket{\psi})^{2}= \rho_{1,1}\cos^{2}\theta+ \rho_{2n,2n} \sin^{2}\theta+ \textup{Re}(\rho_{1,2n}) \sin2\theta
\end{align}

%We can improve this result by noting that we can even use GHZ-states of unequal weight for $\rho'$. Note that in Eq. (\ref{Xstate}) if $b_{i}=0$ for all $i>1$, our state reduces to a GHZ-state of unequal weight. $\hat{X}_{N}=\ket{\psi}\bra{\psi}$
%\begin{align}
%\ket{\psi}=\cos\theta \ket{0,0,,\cdots, 0}+ \sin\theta \ket{1,1, \cdots, 1}
%\end{align}
%Now if we choose $\rho'=\ket{\psi}\bra{\psi}$, the lower bound of the entanglement then reads
%\begin{align}
%\mathcal{E}(\rho)= \min_{\theta} \left(\frac{1}{2}|\sin 2\theta|-\sqrt{1-\bra{\psi}\rho\ket{\psi}^{2}} \right)
%\end{align}
%where 
%\begin{align}
%F(\rho,\ket{\psi})= \rho_{1,1}\cos^{2}\theta+ \rho_{2n,2n} \sin^{2}\theta+ Re(\rho_{1,2n}) \sin2\theta
%\end{align}

%We demonstrate the utility of our approach by using the fidelities recently reported on the creation of GHZ-states among up to fourteen ions to establish lower bounds on the actual entanglement produced in the experiment \cite{PhysRevLett.106.130506}. The lower bounds on the generated all-party entanglement for states that are produced with up to six ions are given in Table \ref{tab:label}. The fidelities of states with larger numbers of qubits are not big enough to establish non-trivial lower bounds. Regardless, these states are all-party entangled as reported by the authors. 

\section{Conclusion}
More than a century after the seminal work of Schmidt \cite{schmidt}, and in spite of important contributions from many, including Werner \cite{PhysRevA.40.4277} and Wootters \cite{wootters}, to the theory of entanglement, the question of how to determine and quantify the entanglement of a given state remains open. Of particular importance is determining the all-party entanglement among many qubits in an experimental setting. This is a difficult problem because most presently available techniques are not well suited for experimentally produced systems in which noise is always present. Additionally, the exponential scaling of measurements required to determine the state of N-qubits make state-based approaches very inefficient. Finally, witness-based techniques for quantifying the entanglement do not currently provide upper bounds and may not result in non-trivial lower bounds.  

In summary, We have introduced a technique to solve this problem by using a distance measure to derive bounds on the all-party entanglement for an arbitrary number N of qubits. We derived an algebraic formula for the all-party entanglement of GHZ-diagonal states and then used this formula to derive easily calculable upper and lower bounds to the all-party entanglement of any experimentally produced state based on the results of just four measurements. This is a particularly promising technique for establishing the bounds of the produced entanglement when N is large. 

As mentioned in the beginning, we used our lower bound to quantify the non-zero all-party entanglement that was produced in a recent experiment \cite{PhysRevLett.106.130506}. We used the fidelities reported on the creation of GHZ-states of up to fourteen ions and established lower bounds on the actual all-party entanglement for states with up to six ions. The fidelities of states with larger numbers of qubits are not big enough to establish non-trivial lower bounds. Regardless, these states are all-party entangled as reported by the authors. While upper bounds appropriate to the experiment in \cite{PhysRevLett.106.130506} could also be computed from two of the populations and one of the coherences using our approach, the required populations were not reported.

\section{acknowledgements}
 We acknowledge partial financial support from ARO W911NF-09-1-0385 and NSF PHY-1203931.

\section{Appendix}

Here we prove that $\chi$, the mapping that takes any density matrix to its X-part, is an LOCC mapping. An algebraic characterization of X-states was presented by \citet{PhysRevA.82.032336}. They have shown that any N-qubit X-state can be written in the following form:
\begin{align}
X=\frac{1}{2^N}\sum_{i=0}^{2^N-1}\left(s_i \hat{S}_i+r_i \hat{R}_i\right),
\end{align}
where $s_0=1$, and $s_i,r_i$ are real. To define $\hat{S}_i$ and $\hat{R}_i$, first note that any generator of SU($2^N$) is a direct product of $N$ generators of SU($2$). $\hat{S}_i$ is the operator obtained by replacing, in the binary representation of $i$, 0's with $\hat{I}$ and 1's with $\hat{\sigma}_z$. $\hat{R}_i$ is the operator that is obtained similarly, but by replacing 0's with $\hat{\sigma}_x$ and 1's with $\hat{\sigma}_y$. For example,
\begin{align}\nn
&\hat{S}_2=\hat{\mathds{1}}~\otimes~ \hat{\mathds{1}}~\otimes \cdots \otimes \hat{\sigma}_z \otimes \hat{\mathds{1}}\\
&\hat{R}_3=\underbrace{\hat{\sigma}_x\otimes \hat{\sigma}_x\otimes \cdots \otimes \hat{\sigma}_y \otimes \hat{\sigma}_y}_\text{$N$}.
\end{align}
Let us define the sets $S=\bigcup_{i=0}^{2^N-1}\{\hat{S}_i\}$ and $R=\bigcup_{i=0}^{2^N-1}\{\hat{R}_i\}$. The set $S\cup R$ is closed under multiplication as well as commutation \cite{PhysRevA.82.032336}. Furthermore, in $S$ there are $2^{N-1}$ operators whose commutation with every element of $S\cup R$ vanishes. These are elements of $S$ with an even number of $\hat{\sigma}_z$'s in their multiplication. We name this subset $C$. Then the quantum operation $\chi $, that takes each N-qubit density matrix and returns its X-part, is given by
\begin{align}
\chi(\rho)=\frac{1}{2^{N-1}}\sum_{\hat{S}_i \in C } \hat{S}_i \rho \hat{S}_i^{\dagger}.
\label{chimapping}
\end{align}
It immediately follows that for any entanglement monotone, the entanglement of $\chi(\rho)$ is always smaller or equal to the entanglement of $\rho$.  The lower bound of all-party concurrence that was derived by \citet{PhysRevA.83.062325} can thus be understood as a special case of this inequality.

Below we prove that the mapping given in Eq. (\ref{chimapping}) returns the X-part of any density matrix. We do so by showing that every generator of SU($2^{N}$), which is not in $S \cup R$ commutes with half of the elements of C and anticommutes with the other half. Since every N-qubit density matrix can be written as a linear sum of generators of SU($2^{N}$), the coefficients of generators in $S \cup R$ survive the sum while the coefficients of the generators outside $S \cup R$ cancel out, leaving only the X-part of the density matrix. The special case of $N=2$ is discussed in \cite{Rau2009}.

Let us first introduce two sets $\xi=\{\mathds{1},\hat{\sigma}_{z}\}$, $\eta=\{\hat{\sigma}_{x},\hat{\sigma}_{y}\}$. Every generator of SU($2^{N}$) is given by a direct product of $N$ operators from either $\eta$ or $\xi$. We consider the arbitrary generator 
\begin{align}
\hat{\mathbf{A}}=\bigotimes_{j=1}^{N} \hat{a}_{j}
\end{align}
where $\hat{a}_{j}$ is either in $\xi$ or $\eta$. We also choose an arbitrary element in C,
\begin{align}
\hat{\mathbf{B}}=\bigotimes_{j=1}^{N} \hat{b}_{j}
\end{align}
where $\hat{b}_{j}$ is either $\hat{\sigma}_{z}$ or $\mathds{1}$, and there are an even number of $\hat{b}_{j}$ for which $\hat{b}_{j}=\hat{\sigma}_{z}$. It is straightforward to show that the commutator has the form
\begin{align}
\hat{\mathbf{A}}\hat{\mathbf{B}}\pm \hat{\mathbf{B}}\hat{\mathbf{A}}=(\bigotimes_{j=1}^{N} \hat{a}_{j}\hat{b}_{j})\pm (\bigotimes_{j=1}^{N} \hat{b}_{j}\hat{a}_{j}).
\end{align}
We note that $\hat{b}_{j}\hat{a}_{j}=-\hat{a}_{j}\hat{b}_{j}$ only if $\hat{a}_{j} \in \eta$ and $\hat{b}_{j}=\hat{\sigma}_{z}$ and $\hat{b}_{j}\hat{a}_{j}=\hat{a}_{j}\hat{b}_{j}$ otherwise. Consequently we find that 
\begin{align}
\hat{\mathbf{A}}\hat{\mathbf{B}}\pm \hat{\mathbf{B}}\hat{\mathbf{A}}=\hat{\mathbf{A}}\hat{\mathbf{B}}(1\pm (-1)^{q})
\end{align}
where $q$ is the total number of times $\hat{a}_{j}\in \eta$ and $\hat{b}_{j}=\hat{\sigma}_{z}$. Thus $\hat{\mathbf{A}}$ and $\hat{\mathbf{B}}$ commute if $q$ is even and they anti-commute otherwise. Therefore we need only count the number of operators in $C$ for which $q$ is even for an arbitrary operator $\hat{\mathbf{A}}$. This is simpler than it seems. Let us assume that there are a total of $M$ $\hat{a}_{j}\in \eta$. The total number of ways an even number of $ \hat{\sigma}_z$'s can be distributed among two sets, one of size M and one of size N-M, with an even number in each is given by
\begin{align}
\sum_{i,j=0} \binom{M}{2i} \binom{N-M}{2j}=2^{N-2}, 
\end{align}
including the case of zero $\sigma_z$'s which represents the identity. The set $C$ has $2^{N-1}$ elements and thus  $\hat{\mathbf{A}}$ anti-commutes with the other half of $C$. Note that this breaks if $M=0$ or $N=0$. But in that case $\hat{\mathbf{A}} \in S\cup R$ and all elements of $C$ commute with $\hat{\mathbf{A}}$. This proves that Eq.~\eqref{chimapping} is an LOCC mapping. 

\bibliographystyle{apsrev4-1}
\bibliography{mybib}

\end{document}